\pdfoutput=1
\documentclass[11pt]{article}

\usepackage[final]{coling}

\usepackage{times}
\usepackage{latexsym}

\usepackage[T1]{fontenc}
\usepackage[utf8]{inputenc}

\usepackage{microtype}

\usepackage{inconsolata}

\usepackage{graphicx}
\usepackage{subfig}

\newcommand{\Set}[1]{\mathcal{#1}}

\newcommand{\Loss}{\mathcal{L}}

\title{Towards Boosting LLMs-driven Relevance Modeling with Progressive Retrieved Behavior-augmented Prompting}

\author{
 \textbf{Zeyuan Chen\textsuperscript{1}},
 \textbf{Haiyan Wu\textsuperscript{2}},
 \textbf{Kaixin Wu\textsuperscript{1}},
 \textbf{Wei Chen\textsuperscript{1}},
 \textbf{Mingjie Zhong\textsuperscript{1}},
 \textbf{Jia Xu\textsuperscript{1}},
\\
 \textbf{Zhongyi Liu\textsuperscript{1}},
 \textbf{Wei Zhang \textsuperscript{3}}
\\
 \textsuperscript{1}Ant Group,
 \textsuperscript{2}Alibaba Group,
 \\
 \textsuperscript{3}School of Computer Science and Technology, East China Normal University
\\
 \small{
   \{chenzeyuan, daniel.wkx, qianmu.cw, mingjie.zmj, steve.xuj, zhongyi.lzy\}@antgroup.com, 
}
\\
 \small{
   wuhaiyan.why@taobao.com, zhangwei.thu2011@gmail.com
 }
}

\begin{document}
\maketitle
\begin{abstract}
\footnotetext{Correspondence to Jia Xu and Wei Zhang.}Relevance modeling is a critical component for enhancing user experience in search engines, with the primary objective of identifying items that align with users' queries. Traditional models only rely on the semantic congruence between queries and items to ascertain relevance. However, this approach represents merely one aspect of the relevance judgement, and is insufficient in isolation. Even powerful Large Language Models (LLMs) still cannot accurately judge the relevance of a query and an item from a semantic perspective. To augment LLMs-driven relevance modeling, this study proposes leveraging user interactions recorded in search logs to yield insights into users' implicit search intentions. The challenge lies in the effective prompting of LLMs to capture dynamic search intentions, which poses several obstacles in real-world relevance scenarios, i.e., the absence of domain-specific knowledge, the inadequacy of an isolated prompt, and the prohibitive costs associated with deploying LLMs. In response, we propose $\textit{ProRBP}$, a novel $\underline{Pro}$gressive $\underline{R}$etrieved $\underline{B}$ehavior-augmented $\underline{P}$rompting framework for integrating search scenario-oriented knowledge with LLMs effectively. Specifically, we perform the user-driven behavior neighbors retrieval from the daily search logs to obtain domain-specific knowledge in time, retrieving candidates that users consider to meet their expectations. Then, we guide LLMs for relevance modeling by employing advanced prompting techniques that progressively improve the outputs of the LLMs, followed by a progressive aggregation with comprehensive consideration of diverse aspects. For online serving, we have developed an industrial application framework tailored for the deployment of LLMs in relevance modeling. Experiments on real-world industry data and online A/B testing demonstrate our proposal achieves promising performance.
\end{abstract}

\section{Introduction}
In today's landscape of excessive information, search engines have become critical for online content platforms, allowing users to swiftly find preferred content that match their search queries. To ensure a user-friendly experience, relevance modeling is crucial to preserve a satisfactory connection between a query and the displayed outcomes, forming a core element of search engine functionality.

In the relevant literature, foundational studies~\citep{robertson2009probabilistic,shah2010evaluating,svore2009machine} have engaged in feature engineering to accomplish text matching, yet they lacked sufficient generalization and accuracy. Subsequently, deep learning-based approaches have risen as a new paradigm, with two primary categories: \textit{representation-based} approaches~\citep{shen2014learning,palangi2014semantic,rao2019bridging} and \textit{interaction-based} approaches~\citep{parikh2016decomposable,chen2016enhanced,hu2014convolutional,pang2016text}. Lately, pre-trained architectures such as BERT~\citep{devlin2018bert} have achieved significant progress in Natural Language Understanding (NLU) tasks. As a result, several studies~\citep{yao2022reprbert,lu2020twinbert,reimers2019sentence,jin2023msra} are introduced that aim to capture the semantic relationships between queries and items. Most recently, Large Language Models (LLMs) have showcased their exceptional capabilities across a wide range of Natural Language Processing (NLP) applications. These models, such as GPT~\citep{radford2019language}, LLaMA~\citep{touvron2023llama}, and GLM~\citep{du2022glm}, are trained on massive corpora of texts, which enables them to maintain an exhaustive world knowledge. Nonetheless, identifying user search intentions accurately remains challenging when relying solely on semantic understanding, due to the absence of specialized domain knowledge required for complex industrial search scenarios. The texts of queries and items in Alipay search scenario are quite short and ambiguous, making it hard to convey effective information contained in their identity. For example, given a query ``Zhe Yi'', the abbreviation of a hospital, it is hard to comprehend the actual semantics. But its historical clicked items include ``the first affiliated hospital of Zhejiang University'', indicating strong correlations between them to help search intention identifying. As such, leveraging behavior data to assist relevance modeling is a natural strategy. 

Existing studies~\citep{chen2023beyond,zeng2022graph,zhu2021textgnn,li2021adsgnn,pang2022improving} have primarily conducted the use of user behavior data. But they all consider constructing the pre-training dataset or the topology structure based on click behaviors without integrating semantics fully and effectively. Despite the success of LLMs, it's still uncertain how well they can integrate world knowledge and specialized domain knowledge represented by user behavior data to master relevance modeling. To this end, this paper aims to investigate the potential of LLMs in relevance modeling with user search behavior. By utilizing strategic prompting techniques, specialized domain knowledge could be easily injected into LLMs for relevance modeling, whereas the performance varies due to the following unresolved issues: (i) \textit{The acquirement of domain-specific knowledge}. Though domain-specific knowledge is vital in improving search scenario-oriented relevance modeling capabilities of LLMs, not all knowledge is beneficial. The noisy user behavior data may mislead LLMs to undesired judgements. Moreover, specialized domain knowledge of search scenarios undergoes rapid changes on a daily basis. The limited capacity of LLMs to adapt swiftly to these changes presents a significant obstacle to their ability to render accurate relevance judgments. (ii) \textit{The inadequacy of an isolated prompt}. Despite LLMs could derive the relevance degree exploiting an isolated prompt, LLMs exhibit insensitivity to input, meaning they lack awareness of the aspects from which to infer relevance. In addition, the isolated prompts place greater demands on the quality of the prompts itself. Except for the aforementioned issues, deploying LLMs affordably in industrial scenarios is also a consideration worth addressing.

To address the above problems, we propose a novel $\underline{Pro}$gressive $\underline{R}$etrieved $\underline{B}$ehavior-augmented $\underline{P}$rompting framework for integrating search scenario-oriented knowledge with LLMs (dubbed as $\textit{ProRBP}$). To acquire domain-specific knowledge in time, we perform a user-driven behavior neighbor retrieval from the daily updated search logs, retrieving candidates that users consider to meet their expectations currently. Then we anticipate employing advanced prompting techniques that progressively improve the outputs of the LLMs, followed by a progressive aggregation with comprehensive consideration of diverse aspects to form a holistic relevance model. As for the online serving of LLMs, we design an industrial implementation framework enabling LLMs to fully handle search relevance scenarios with the affordable cost.

In summary, we make the following contributions of this paper:
\begin{itemize}
    \item To the best of our knowledge, we are the first to successfully investigate the potential of LLMs with user behavior data to master relevance modeling.
    
    \item We propose $\textit{ProRBP}$ with two novel plug-in modules. Firstly, a user-driven behavior neighbors retrieval is developed to acquire domain-specific knowledge in time.
    Secondly, the proposal of progressive prompting and aggregation can strengthen the judgement of relevance for LLMs.

    \item We explore an industrial implementation~\ref{sec:ii} enabling LLMs to fully handle search relevance scenarios in Alipay search with the affordable cost and latency.

    \item We demonstrate $\textit{ProRBP}$ framework achieves superior performance through experiments on real-world industry data and online A/B testing. It has been deployed online and outperforms prior approaches~\citep{chen2023beyond} on core metrics.
\end{itemize}

\section{Related Work}\label{sec:related}
Relevance modeling in search can be viewed as a text matching problem as the sub-domain of information retrieval (IR). The majority of work focuses on distinctions at the semantic level, while a minority of methods judge the relevance from a behavioral perspective.

Current semantics-driven approaches can be classified into two aspects: \textit{feature-based} approaches and \textit{deep learning-based} approaches.
The first category is centered on manual-crafted features such as TF-IDF similarity and BM25~\citep{svore2009machine}. Despite their usefulness, these feature-based approaches have limited generalization ability due to their domain-specific features and require significant labor resources. 

In order to address the limitations of the above approaches, deep learning-based approaches emerge as the new paradigm, which can be broadly classified into \textit{representation-based} approaches and \textit{interaction-based} approaches. The former focuses on learning a low-dimensional representation of data while the latter emphasizes capturing the interaction between inputs. For instance, DSSM~\citep{shen2014learning} is a classical two-tower representation-based model that encodes the query and the document separately. In this paradigm, recurrent~\citep{palangi2014semantic,tai2015improved} and convolutional~\citep{hu2014convolutional,shen2014learning} networks are adopted to extract low-dimensional semantic representations. For these methods, the encoding of each input is carried out independently of the others. Consequently, these models face challenges in modeling complex relationships. To overcome this limitation, interaction-based models are proposed. DecompAtt~\citep{parikh2016decomposable} leverages attention network to align and aggregate representations. In parallel, recurrent~\citep{chen2016enhanced} and convolutional~\citep{hu2014convolutional,pang2016text} networks are employed for modeling complex interactions.

In recent times, pre-trained models like BERT~\citep{devlin2018bert} have made remarkable strides in Natural Language Understanding (NLU). As a result, representation-based~\citep{yao2022reprbert,lu2020twinbert,reimers2019sentence,jin2023msra} and interaction-based architectures~\citep{wang2019structbert} are proposed to leverage the capabilities of these models to encode semantic correlations between queries and items. Most recently, Large Languages Models (LLMs) like GPT~\citep{radford2019language}, LLaMA~\citep{touvron2023llama}, BLOOM~\citep{workshop2022bloom} and GLM~\citep{du2022glm} trained on massive corpora of texts have shown their superior ability in language understanding, generation, interaction, and
reasoning tasks. ~\citep{sun2023chatgpt} investigates the potential of utilizing LLMs for searching and demonstrates that appropriately instructs ChatGPT and GPT-4 can produce competitive and even superior results to supervised methods widely used information retrieval benchmarks. ~\citep{chen2023beyond} tries to deal with long-tail query-item matching through LLMs efficiently and effectively. In these research work, they tend to exploit world knowledge stored in parameters of LLMs to judge the relevance between the query and item. However, general LLMs can not adapt to the industrial scenario due to the lack of domain-specific knowledge and insensitivity to the relevance judgement. In this work, we try to explore the LLMs-driven relevance modeling comprehensively in Alipay search engine to meet the above requirements.

In addition to textual information, there are a few related works that aim to integrate user behavior data into their models. The utilization of user behavior data can provide valuable insights into search intention, which can then be used to enhance the relevance of the search engines. MASM~\citep{yao2021learning} leverages the historical behavior data to complete model pre-training as a weak-supervision signal with a newly proposed training objective. ~\citep{zhu2021textgnn,li2021adsgnn,pang2022improving} try the incorporation of click graphs to enhance the effectiveness of search systems. ~\citep{chen2023beyond} endeavors to exploit behavior neighbors while considering interaction granularity and topology structure. However, no work has fully integrated LLMs with user behavior comprehensively in relevance modeling. We target to do this.

\section{Problem Formulation}\label{sec:pf}
Assume we have the target query $q$ and target item $i$ needed to predict the relevance degree exploiting LLMs. In essence, referring to PET~\citep{schick2020exploiting}, it could be formulated as follows: with the designed prompt $\tau(q,i)$, LLMs can determine which verbalizer $v$ (i.e., ``relevant'' or ``irrelevant'') is the most likely substitute for the mask based on the likelihood $\mathcal{P}(v|\tau(q,i))$. The most naive prompt in the relevance modeling could be formulated as:
\begin{equation}
    \tau(q,i) = Is\,[q]\,and\,[i]\,related?\,[mask]\,, 
\end{equation}
The relevance label $y_{qi}\in\{0,1\}$ can be associated with a verbalizer (i.e., ``irrelevant'' or ``relevant'') from the vocabulary of LLMs to denote the relevance degree between $q$ and $i$. To enable the adaptation of general LLMs to the relevance modeling task, supervised fine-tuning operation is selected using the cross-entropy loss function $\mathcal{L}_{ce}$. Then the relevance degree could be given from LLMs for subsequent applications in Alipay search scenario. It worth noting that the above formulation is the basic explanation of relevance modeling using LLMs. We will further explore novel ways below based on this basic.

\section{Methodology}
\begin{figure*}[!t]
    \centering
	\includegraphics[width=0.85\linewidth]{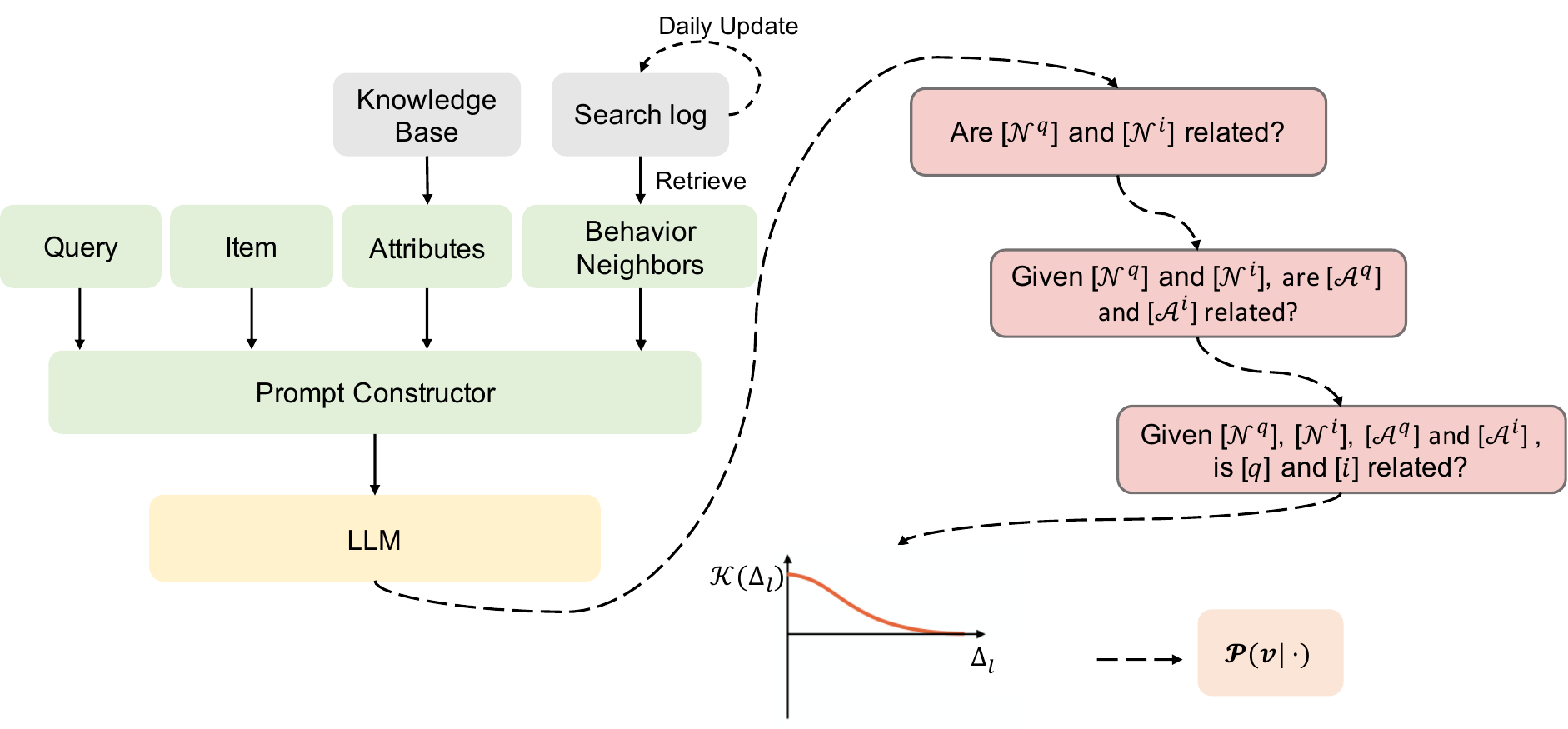}
    % \vspace{-0.5em}
    \caption{The proposed framework $\textit{ProRBP}$.}
    \label{fig:architecture}
    % \vspace{-1.em}
\end{figure*}

In this section, we elaborate the proposed $\textit{ProRBP}$ framework with its novel plug-in modules as depicted in Figure~\ref{fig:architecture}.

\begin{figure}[!t]
    \centering
	\includegraphics[width=0.91\linewidth]{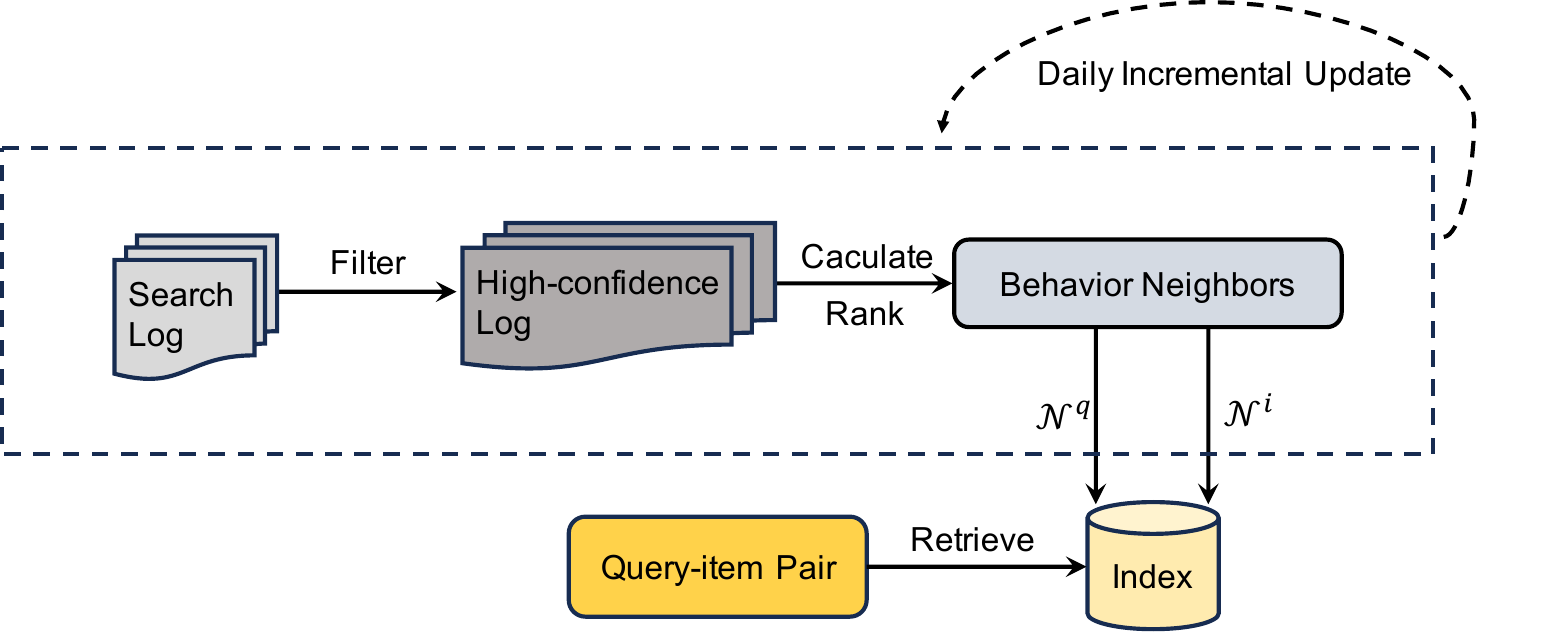}
    % \vspace{-1.5em}
    \caption{The pipeline of user-driven behavior neighbor retrieval.}
    \label{fig:BNR}
    % \vspace{-1.5em}
\end{figure}

\subsection{User-driven Behavior Neighbor Retrieval}
Generally, LLMs possess an exhaustive world knowledge with the benefit of massive corpora of texts pre-training. However, LLMs still struggle to understand user search intentions for short and ambiguous queries and items from Alipay search due to the lack of specialized and rapidly evolving domain knowledge. A promising approach for addressing the above is retrieval-augmented language modeling~\citep{gao2023retrieval}, grounding the LLMs during generation by conditioning on relevant candidates retrieved from an external knowledge source. Drawing inspiration from this, we devise a user-driven behavior neighbor retrieval module. This module could retrieve the daily search logs of users to obtain daily changing behavior neighbors that users consider relevant. The pipeline of user-driven behavior neighbor retrieval is depicted as Figure~\ref{fig:BNR}.
% Though this module could address the mentioned problem efficiently, not all knowledge is beneficial.
Due to the limited number of items displayed for a query in Alipay search scenario, almost all results can be seen by users. With this in mind, we could analyze that a higher click-through rate reflects that users believe the corresponding query-item pairs can better meet their search intents and needs when the exposure PV (i.e., page view) reaches a certain quantity. Thus, we filter out the query-item pairs with less than 100 exposure PV and formulate the remaining search logs as high-confidence logs. Then, we utilize the high-confidence logs from the past month to calculate the click-through rate for the exposed query-item pairs. To mitigate the effect of noises, query-item pairs are selected above a click-through threshold (e.g., 0.2) and the neighbors are arranged in descending order based on click-through rate from the query and item perspective, respectively. And we only choose top-K (e.g., 20) neighbors for corresponding query and item. Through this approach, it is possible to select dual behavior neighbors with high confidence, which in turn can assist LLMs in the process of in-context learning~\citep{ram2023context,wang2023boosting,liao2024d2llm,liu2024boosting,wei2024llmgr}. To ensure the timeliness of knowledge, we utilize the daily updated search logs to repeat the above operation every day and construct daily separate indexes for the dual behavioral neighbors of queries and items. The neighbors could be retrieved from indexes for the corresponding query $q$ and item $i$, denoted as $\Set{N}^q$ and $\Set{N}^i$ respectively. Need to clarify is that though we can obtain the behavior neighbors for the majority of queries or items, there is still a small portion of queries or items whose behavior neighbors cannot be obtained. In response to this situation, we will set them as empty in the prompt. Besides, we could obtain the attributes (i.e., brand, keyword, intent) of the query $q$ and item $i$ from the industrial knowledge base, which could be denoted as $\Set{A}^q$ and $\Set{A}^i$. They can be treated as a supplement to the domain-specific information. Then we could construct our daily prompt $\tau(q,i,\Set{A}^q,\Set{A}^i,\Set{N}^q,\Set{N}^i)$ based on the above information. The daily prompt will be send to LLMs for further operation.

\subsection{Progressive Prompting and Aggregation}
Despite the fact that LLMs can determine a relevance score using an isolated prompt, they exhibit insensitivity to the input. It indicates that they lack the awareness required to decide the aspects from which to infer relevance. Furthermore, relying on isolated prompts imposes higher demands on the quality of the prompts itself, as the diversity of domain knowledge and the sensitivity to slight modifications of prompts may lead to unexpected results. As a result, it's expected that using progressive prompting and aggregation process to gradually guide and consider diverse aspects will eventually lead to the creation of a unified model that assesses relevance.

\textbf{Progressive Prompting.} It is designed for making LLMs sensitive to the diverse aspects for the relevance judgement and improving the robustness of the model performance. Specifically, we firstly decompose the mentioned prompt above and construct least-to-most prompts step by step~\citep{zhou2022least,wei2022chain} as:
{\footnotesize
\begin{equation}
\tau(\Set{N}^q,\Set{N}^i)\hookrightarrow \tau(\Set{A}^q,\Set{A}^i,\Set{N}^q,\Set{N}^i)\hookrightarrow \tau(q,i,\Set{A}^q,\Set{A}^i,\Set{N}^q,\Set{N}^i)\,, 
\end{equation}
}
The prompts are documented as shown in Figure~\ref{fig:architecture}. Besides, different prompting documents could obtain the consistent improvement in our local experiments exploiting this module.

Then LLMs could reason the likelihood of the verbalizer exploiting the least-to-most prompts step by step for the sensitivity to relevance judgement and stable prediction results.
% , which could be dubbed as:
% {\tiny
% \begin{equation}
%     \mathcal{P}(v|\tau(\Set{N}^q,\Set{N}^i))\hookrightarrow \mathcal{P}(v|\tau(\Set{A}^q,\Set{A}^i,\Set{N}^q,\Set{N}^i))\hookrightarrow \mathcal{P}(v|\tau(q,i,\Set{A}^q,\Set{A}^i,\Set{N}^q,\Set{N}^i))\,, 
% \end{equation}
% }
Simply, all least-to-most prompts share the same relevance label $y_{qi}$ and all least-to-most sub-tasks are supervised by the cross-entropy loss in parallel explicitly. The aforementioned method can be extended to $L$ progressive prompts and the supervised loss could be denoted as $\Loss_{auxi}= \sum_{l=1}^L \Loss_{ce}^l$, where $\Loss_{ce}^l$ represents the loss of the sub-task corresponding to $l$-th prompt.

\textbf{Progressive Aggregation.} Once we obtain the $L$ least-to-most probabilities, we expect to learn the progressive relationship of solutions to sub-tasks. Intuitively, Our least-to-most paradigm is incremental not only in terms of information volume, but also in importance. Hence, we adopt the kernel function $\mathcal{K}$ to model the incremental tendency and the choice of kernel function can be a Gaussian kernel function, exponential kernel function, logarithmic decay kernel function, etc. After experimental comparison, we have chosen the exponential kernel function $\mathcal{K}(\Delta_l) = Exp(\Delta_l|\lambda)$, where $\lambda$ is a learnable parameter and $\Delta_l$ is the degree of attenuation of the $l$-th prompt defined by us. The overall relevance score could be acquired through aggregating the probabilities from the least-to-most sub-tasks progressively as:
\begin{equation}
    \Set{P}(v|\cdot) = \sum_{l=1}^L\mathcal{K}(\Delta_l)\times \mathcal{P}(v|\tau_l)\,, 
\end{equation}
This relevance score could be treated as the production of overall task, deserving to supervise mainly. The loss function is given by $\Loss_{main}$. The hybrid objective function can be formulated as:
 \begin{equation}\label{eq:loss}
     \Loss = \Loss_{main}+\alpha\Loss_{auxi}\,, 
 \end{equation}
where $\alpha$ is a hyper-parameter to control the strength of the least-to-most sub-tasks.

\section{Experiments}
% In this section, we conduct extensive experiments to validate the effectiveness of our proposed $\textit{ProRBP}$ framework by answering the following pivotal research questions:
% \begin{itemize}
% \item[\textbf{RQ1.}] How are the results of $\textit{ProRBP}$ framework compared with other competitive relevance modeling models?

% \item[\textbf{RQ2.}] How do the main components of $\textit{ProRBP}$ framework affect its relevance modeling performance?
% \end{itemize}

% Besides, we investigate the effect of different hyper-parameter settings (e.g., the number of behavior neighbors, the type of the kernel functions, and the strength of the least-to-most sub-tasks). Finally, we conduct online A/B testing to show the performance of the proposed method compared to previous deployed model.

\subsection{Experimental Setup}
\textbf{Datasets.} In order to evaluate the performance of all the models with reliability, we select the real-world industry data used in mini apps search scenario of Alipay search engine and present its statistics in Table~\ref{tbl:stat}. The dataset is labeled by human annotators, where Good and Bad annotations denote label 1 and 0 respectively. User historical behavior data is sampled from the search logs of the search engine. Although datasets such as WANDS\footnote{\url{https://github.com/wayfair/WANDS/tree/main}} and MSLR\footnote{\url{https://www.microsoft.com/en-us/research/project/mslr/}} are publicly available, they do not contain the requisite user historical behavior data. Hence, we select this in-house data to evaluate the proposed framework. Other related work~\citep{yao2022reprbert,yao2021learning,zhu2021textgnn,li2021adsgnn,chen2023beyond} also selects one in-house data to evaluate the proposed approaches. The partial data of this paper was released before\footnote{\url{https://github.com/alipay/BehaviorAugmentedRelevanceModel}}. 

\textbf{Baseline Models.} We select a set of popular NLU (Natural Language Understanding)-based, behavior-based and NLG (Natural Language Generation)-based relevance models as baselines. For the NLU-based models, we choose three common models including two-tower and single-tower architectures: DSSM~\citep{shen2014learning}, ReprBert~\citep{yao2022reprbert}, Bert~\citep{devlin2018bert}. The behavior-based models all consider constructing the pre-training dataset (MASM~\citep{yao2021learning}) or the topology structure based on click behaviors (TextGNN~\citep{zhu2021textgnn}, AdsGNN~\citep{li2021adsgnn}, BARL-ASe~\citep{chen2023beyond}). For the NLG-based models, we choose two foundation models including two set of architectures: the causal decoder (BLOOM~\citep{workshop2022bloom}) and the prefix decoder (GLM~\citep{du2022glm}).

\begin{table}[!t]
\centering
\caption{Statistics of the human-annotated dataset.}\label{tbl:stat}
% \vspace{-1.em}
\resizebox{1.0\linewidth}{!}{
\begin{tabular}{cccccc}
\hline
Dataset  & \# Sample  & \# Query  & \# Item  & \# Good & \# Bad\\ \hline
 Train & 773,744 & 87,499 & 88,724 & 460,610 & 313,134\\
 Valid & 97,032 & 40,754 & 25,192 & 57,914 & 39,118\\
 Test & 96,437 & 40,618 & 25,004 & 57,323 & 39,114\\
\hline
\end{tabular}
}
% \vspace{-1.em}
\end{table}

\textbf{Evaluation Metrics.} We use Area Under Curve (AUC), F1-score (F1), and False Negative Rate (FNR) to measure the multidimensional performance of all models. AUC and F1 are commonly used in the studied area, of which higher metric values represent better model performance. Conversely, lower False Negative Rate (FNR) values are preferable, as they indicate a lower false filtering rate of models. Note that AUC often serves as the most significant metric in our task while the others provide auxiliary supports for our analysis.

\textbf{Model Implementations.} We tune our model for 5 epochs with batch size 64 and learning rate 3e-05 in the supervised fine-tuning stage of LLMs. For the foundation models, we select the 1.1B BLOOM\footnote{\url{https://huggingface.co/bigscience/bloom-1b1}} and GLM with different magnitude of parameters (e.g., 0.3B, 2B and 10B) pre-trained by Alipay. The number of retrieved behavior neighbors is set to 20. We tune the parameter $\alpha$ within the ranges of \{0, 0.05, 0.1, 0.2, 0.3, 0.5, 1.0\}. We conduct the experiments on NVIDIA Tesla A100 GPUs.

\begin{table}[!t]
\centering
\caption{Main results on real-world industry data. (-) denotes the lower value corresponds to better performance. Improvements over variants are statistically significant with p < 0.05.}\label{tbl:performance-comp}
% \vspace{-0.3em}
\resizebox{0.98\linewidth}{!}{
\begin{tabular}{c|ccc} % |*{7}{c|}
\hline

\hline
Method
&\multicolumn{1}{c}{AUC} &\multicolumn{1}{c}{F1} &\multicolumn{1}{c}{FNR (-)}  \\ \hline

DSSM
&\multicolumn{1}{c}{0.8356} &\multicolumn{1}{c}{0.8210} &\multicolumn{1}{c}{0.1389}  \\

ReprBert
&\multicolumn{1}{c}{0.8388} &\multicolumn{1}{c}{0.8376} &\multicolumn{1}{c}{0.1280} \\

Bert
&\multicolumn{1}{c}{0.8540} &\multicolumn{1}{c}{0.8534} &\multicolumn{1}{c}{0.1150} \\

MASM
&\multicolumn{1}{c}{0.8547} &\multicolumn{1}{c}{0.8318} &\multicolumn{1}{c}{0.1289} \\

TextGNN
&\multicolumn{1}{c}{0.8847} &\multicolumn{1}{c}{0.8489} &\multicolumn{1}{c}{0.1290} \\

AdsGNN
&\multicolumn{1}{c}{0.8878} &\multicolumn{1}{c}{0.8458} &\multicolumn{1}{c}{0.1454} \\

BARL-ASe
&\multicolumn{1}{c}{0.9078} &\multicolumn{1}{c}{0.8658} &\multicolumn{1}{c}{0.1054}  \\

\hline

GLM-0.3B
&\multicolumn{1}{c}{0.8608} &\multicolumn{1}{c}{0.8585} &\multicolumn{1}{c}{0.1106} \\

\quad\quad+$\textit{ProRBP}$
&\multicolumn{1}{c}{0.9105} &\multicolumn{1}{c}{0.8778} &\multicolumn{1}{c}{0.1035} \\

\hline

BLOOM-1.1B
&\multicolumn{1}{c}{0.8543} &\multicolumn{1}{c}{0.8511} &\multicolumn{1}{c}{0.1174} \\

\quad\quad+$\textit{ProRBP}$
&\multicolumn{1}{c}{0.8991} &\multicolumn{1}{c}{0.8675} &\multicolumn{1}{c}{0.1121} \\

\hline

GLM-2B
&\multicolumn{1}{c}{0.8619} &\multicolumn{1}{c}{0.8598} &\multicolumn{1}{c}{0.1065} \\

\quad\quad+$\textit{ProRBP}$
&\multicolumn{1}{c}{0.9120} &\multicolumn{1}{c}{0.8776} &\multicolumn{1}{c}{0.1007} \\

\hline

GLM-10B
&\multicolumn{1}{c}{0.8751} &\multicolumn{1}{c}{0.8656} &\multicolumn{1}{c}{0.1006} \\

\quad\quad+$\textit{ProRBP}$
&\multicolumn{1}{c}{0.9143} &\multicolumn{1}{c}{0.8801} &\multicolumn{1}{c}{0.0973} \\

\hline

\hline
\end{tabular}
}
% \vspace{-1.em}
\end{table}

\subsection{Experimental Results}
\textbf{Performance Comparison.} Table~\ref{tbl:performance-comp} shows the overall comparison with baselines. Our findings indicate that DSSM performs poorly since it merely encodes the embeddings of the query and item independently. By further comparing DSSM with ReprBert, we find the performance is improved to a certain degree. This demonstrates the pre-trained models utilizing the large corpus can bring additional gains. Compared to ReprBert, Bert achieves better performance, as the interaction-based models possess more advanced text relevance modeling abilities than representation-based models. 

For the models of MASM, TextGNN, and AdsGNN, they exploit historical behavior data to build pre-training dataset or click graph to enhance the effectiveness of search systems. And they achieve significantly better results than the above-mentioned methods in the metric of AUC rather than F1 and FNR. This can be attributed to the introduction of auxiliary signals, which may bring improvement for the ranking ability of models but introduce some noises leading to the loss of F1 and FNR. The performance differences among them depend on the utilization of behavior data and modeling granularity. The newly proposed model BARL-ASe achieves the best performance in behavior-based relevance models.

GLM-0.3B, BLOOM-1.1B, GLM-2B and GLM-10B denote that we use the corresponding language models with the amount of parameters to perform relevance judgements exploiting the naive prompts as stated in Section~\ref{sec:pf}. They exhibit relatively good performance, demonstrating the positive effect of massive training datasets and large-scale parameters. Performance variances among them with different parameters appear minimal, possibly because the simplicity of the relevance task doesn't fully tap into the advantages of LLMs. Their performance has a certain gap compared to behavior-based models especially on the core metric AUC. This may be attributed to the lack of domain-specific knowledge and the limitations of an isolated prompt. Thus, $\textit{ProRBP}$ framework is proposed to address the mentioned problem and significantly improve the foundation models' performance. The performance differences between BLOOM and GLM may stem from differences in architecture and training corpora. 

Overall, exploiting our framework $\textit{ProRBP}$ could yield the best performance and ensures efficiency and economy. In the online serving~\ref{sec:ii}, we select GLM-10B$+\textit{ProRBP}$ to perform offline inference and GLM-2B$+\textit{ProRBP}$ to proceed online prediction considering affordable cost and efficiency. Besides, they both obtain better gains under all the metrics on the evaluation dataset compared to the second-best performed model BARL-ASe that was deployed in our scenario before.
\begin{table}[!h]
\centering
% \vspace{-0.2em}
\caption{Ablation study of GLM-2B+$\textit{ProRBP}$}\label{tbl:abl}
% \vspace{-0.2em}
\resizebox{0.95\linewidth}{!}{
\begin{tabular}{c|ccc} % |*{7}{c|}
\hline
Method
&\multicolumn{1}{c}{AUC} &\multicolumn{1}{c}{F1} &\multicolumn{1}{c}{FNR (-)}  \\
\hline

$\textit{ProRBP}$
&\multicolumn{1}{c}{0.9120} &\multicolumn{1}{c}{0.8776} &\multicolumn{1}{c}{0.1007}  \\

\quad\quad-BNR
&\multicolumn{1}{c}{0.8943} &\multicolumn{1}{c}{0.8656} &\multicolumn{1}{c}{0.1037} \\

\quad\quad-PPA
&\multicolumn{1}{c}{0.8872} &\multicolumn{1}{c}{0.8645} &\multicolumn{1}{c}{0.1045} \\

\quad\quad-Both
&\multicolumn{1}{c}{0.8619} &\multicolumn{1}{c}{0.8598} &\multicolumn{1}{c}{0.1065} \\

\hline
\end{tabular}
}
% \vspace{-0.5em}
\end{table}

\textbf{Ablation Study.} To investigate the contributions of key components, we provide the following variants of our complete framework:
(1) ``-BNR'' denotes discarding user-driven behavior neighbor retrieval strategy;
(2) similarly, ``-PPA'' denotes discarding progressive prompting and aggregation module. 
(3) ``-Both'' represents removing two modules mentioned above.

The results shown in Table~\ref{tbl:abl} lead to the conclusion that both ``BNR'' and ``PPA'' strategies yield significantly positive results. And ``PPA'' demonstrates greater significance than ``BNR''.
% To investigate the contributions of key modules adopted by $\textit{ProRBP}$ framework, we provide the following variants of our complete framework:
% \begin{itemize}
% \item ``-BNR'' denotes discarding user-driven behavior neighbor retrieval module.
% \item ``-PPA'' denotes discarding progressive prompting and aggregation module. 
% \item ``-Both'' represents removing two modules mentioned above simultaneously.
% \end{itemize}

% Throughout the ablation study shown in Table~\ref{tbl:abl}, we observe that:
% \begin{itemize}
%     \item[$\diamond$] By analyzing the results of ``-BNR'' and ``PPA'', we could conclude that both ``BNR'' and ``PPA'' modules yield significantly positive results. This suggests that the importance of the domain-specific knowledge and advanced prompting techniques in LLM-driven relevance modeling. 
    
%     \item[$\diamond$] And ``PPA'' demonstrates greater significance than ``BNR''. It is likely that the naive prompts with domain-specific knowledge can not release the potentials of LLMs totally for relevance modeling. It proves that the improvement of the sensitivity of LLMs to relevance judgement and prediction stability brings more gains than domain-specific knowledge in our task. 
    
%     \item[$\diamond$] The result of ``-Both'' demonstrates removing two modules could significantly degrade the model's performance.
% \end{itemize}

\subsection{Parameter Sensitivity}
\begin{figure}[!h]
    \centering
    % \vspace{-1.em}
    \subfloat
    {
    \includegraphics[width=0.32\linewidth]{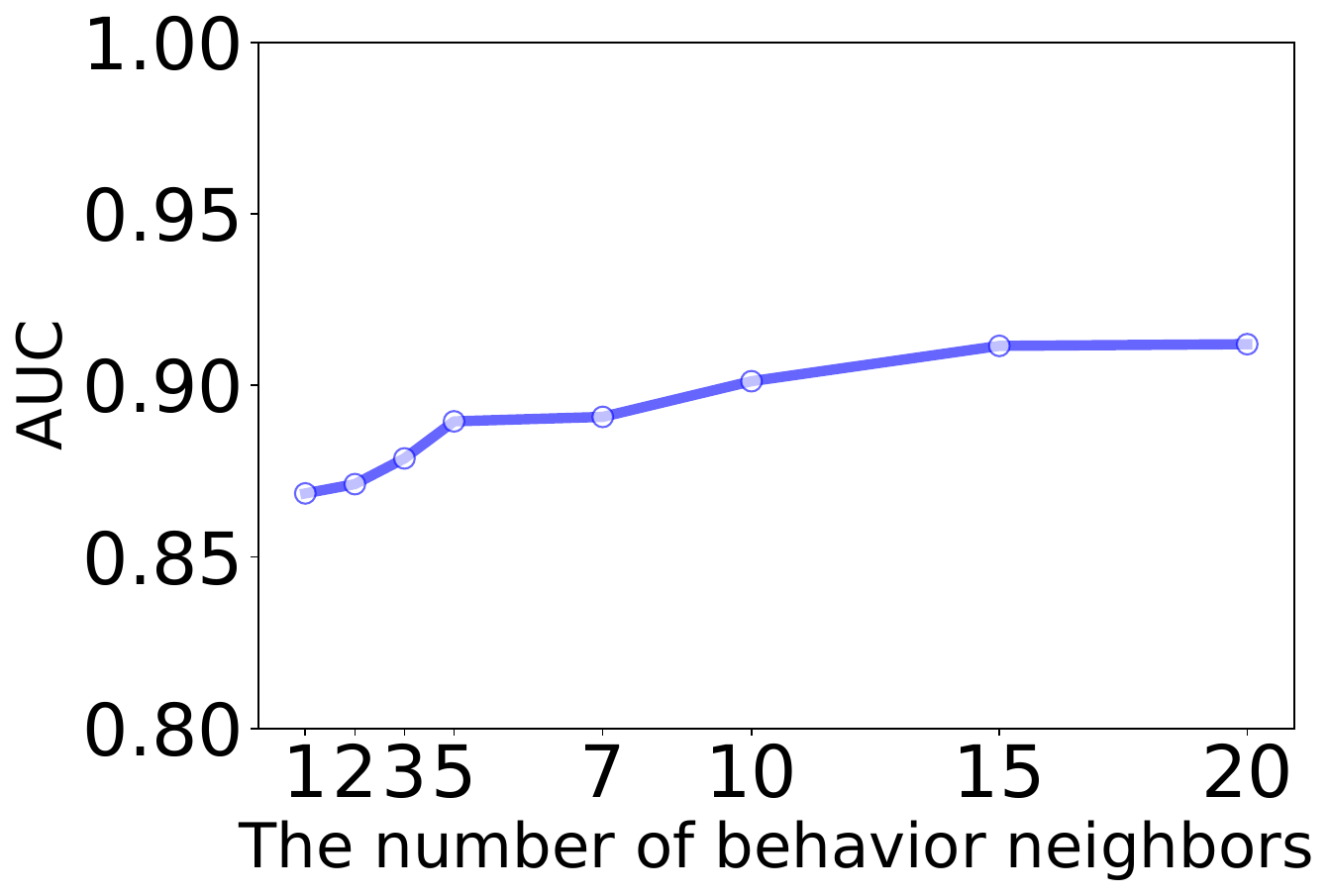}\label{fig:neighbor}
    \includegraphics[width=0.32\linewidth]{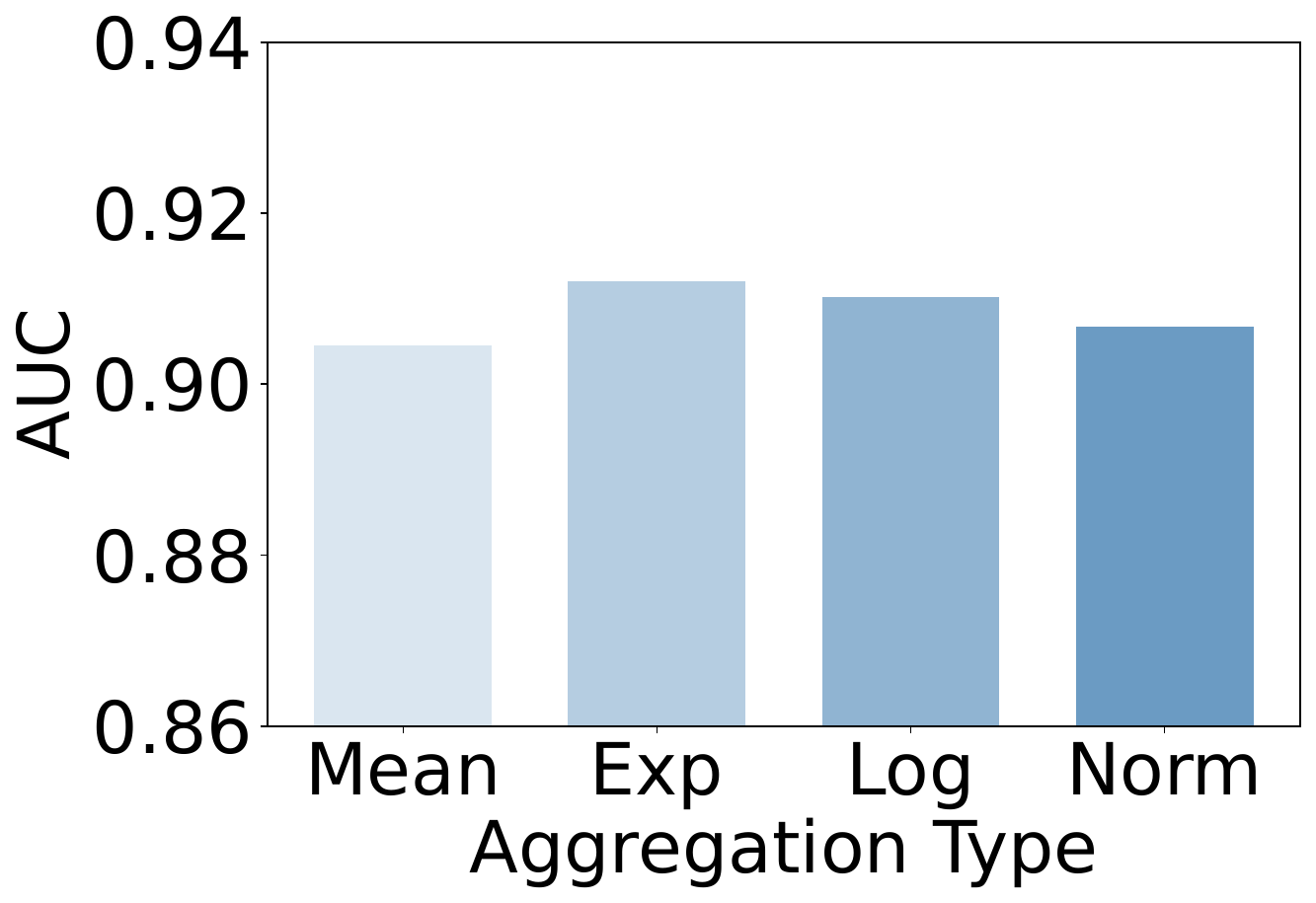}\label{fig:agg}
    \includegraphics[width=0.32\linewidth]{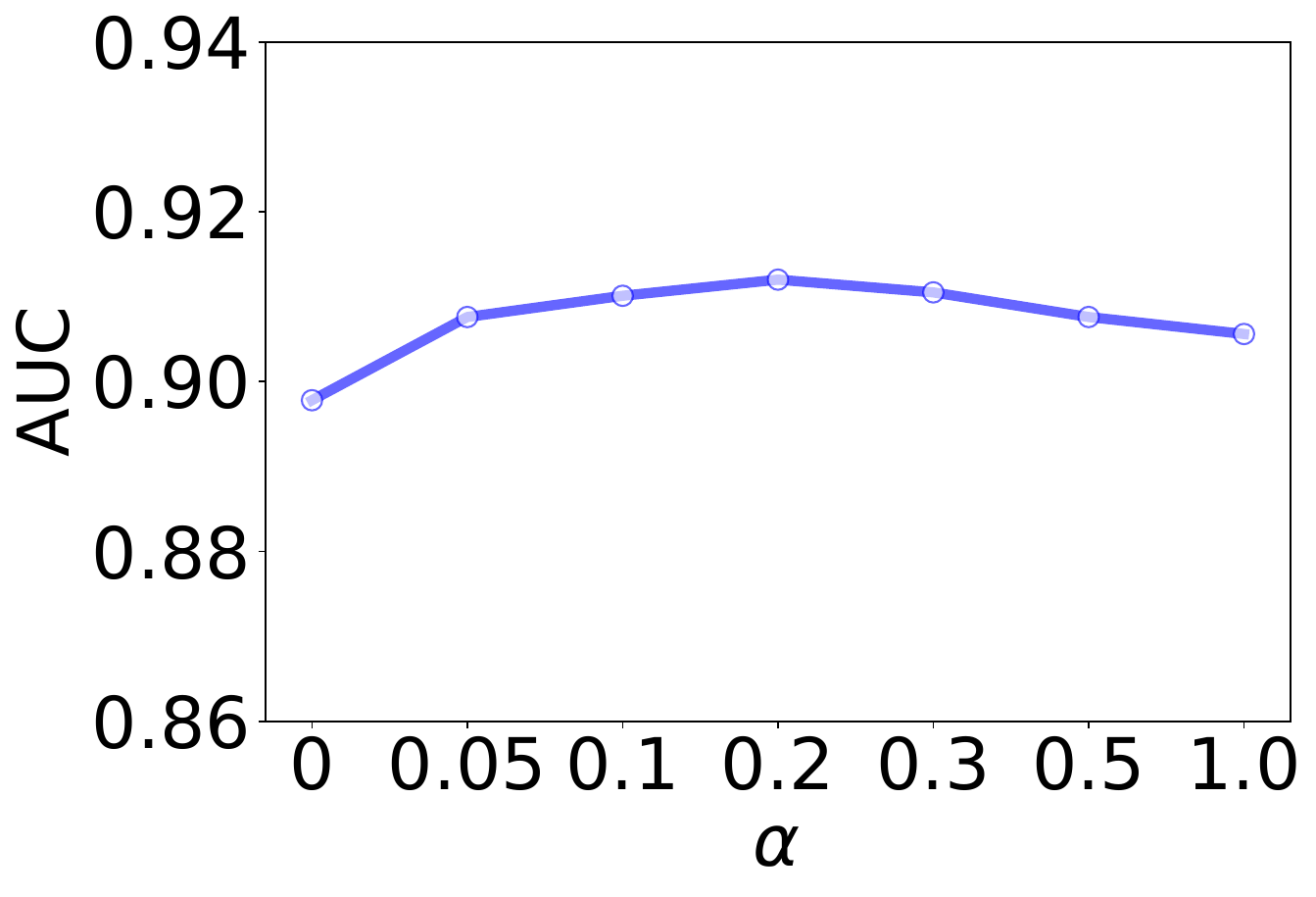}\label{fig:alpha}
    }
    % \vspace{-0.7em}
    \caption{Result variation with different settings.}
    \label{fig:var}
    % \vspace{-0.7em}
\end{figure}

From the variation trends of Figure~\ref{fig:var}, we could observe that: (1) As the number of behavioral neighbors increases, the model's performance continues to improve, tending to plateau at around 20.
(2) The exponential kernel function yields the best result among the mean-pooling (Mean), Gaussian (Norm) and logarithmic (Log) kernel functions. (3) The better results are achieved when setting $\alpha$ to a suitable value. Too large (e.g., 1.0) or small (e.g., 0.01) value will lead to a performance drop.

\subsection{Online A/B Testing}
The proposed method has been deployed in the relevance stage of Alipay search platform providing search service of mini apps and demonstrated its significant performance gains in online A/B testing compared with the previous model BARL-ASe. The each experiment takes about 3\% proportion of Alipay search traffic for two weeks. Compared with the previously deployed model, the proposed method improves the valid PV-CTR\footnote{the number of valid clicks divided by the number of searches} by 0.33\% on average without causing an increase in latency. And the results of human annotations show the model can reduce the rate of irrelevant results by 1.07\% points on average. The results demonstrate that our proposal can improve the experience of users.

\section{Conclusion}
This paper studies the relevance modeling problem by integrating world knowledge stored in the parameters of LLMs with specialized domain knowledge represented by user behavior data for achieving promising performance. The novel framework $\textit{ProRBP}$ is proposed, which innovatively develops user-driven behavior neighbor retrieval module to learn domain-specific knowledge in time and introduces progressive prompting and aggregation module for considering diverse aspects of the relevance and prediction stability. We explore an industrial implementation to deploy LLMs to handle full-scale search traffics of Alipay with acceptable cost and latency. The comprehensive experiments on real-world industry data and online A/B testing validate the superiority of our proposal and the effectiveness of its main modules.

\bibliography{custom}

\appendix

\section{Appendix}\label{sec:appendix}

\subsection{Industrial Implementation for LLMs}\label{sec:ii}
\begin{figure}[!t]
    \centering
	\includegraphics[width=\linewidth]{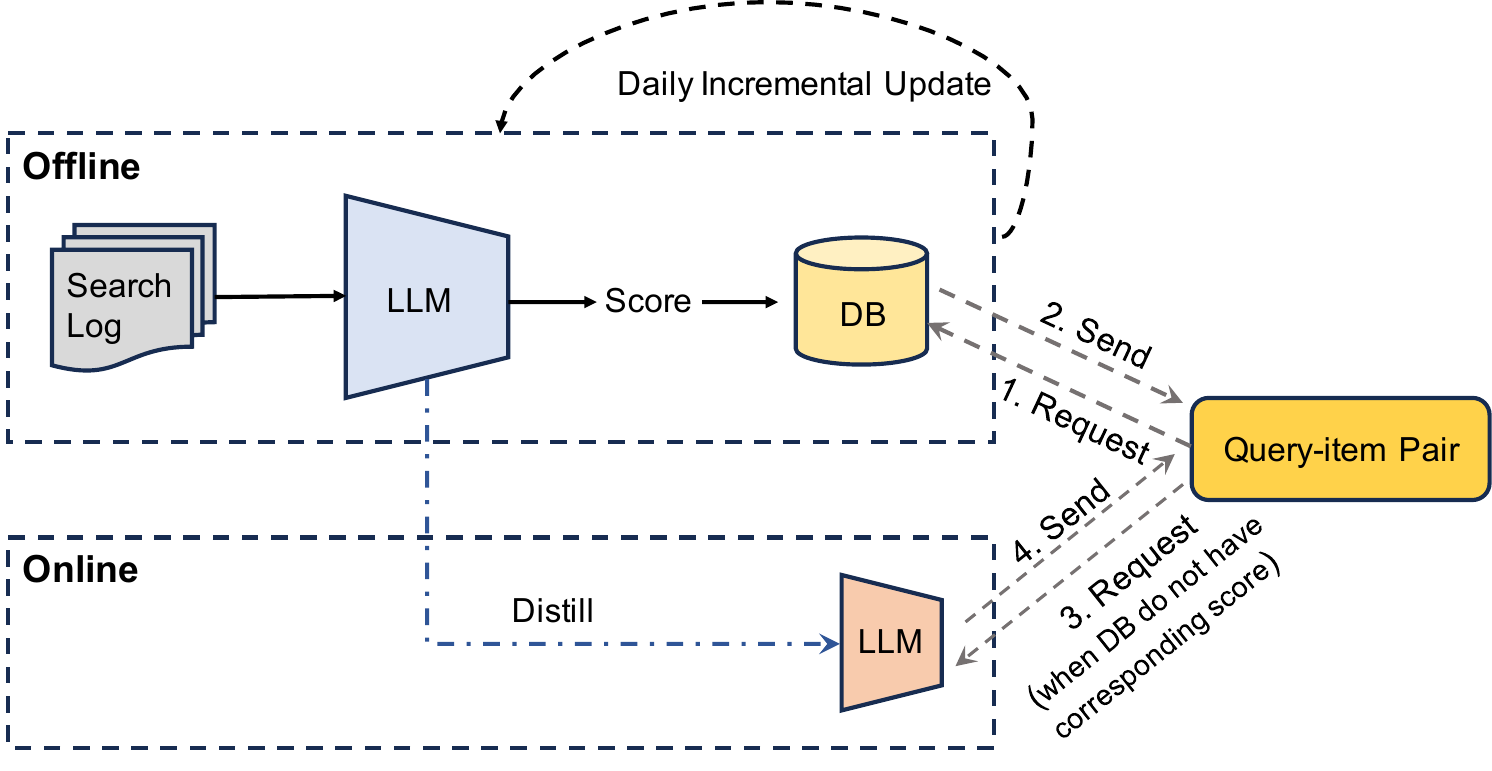}
    % \vspace{-1.5em}
    \caption{The illustration of online and offline collaborative service.}
    \label{fig:OOCS}
    % \vspace{-1.5em}
\end{figure}

In real-world search scenarios (e.g., Alipay search), it is hard to deploy LLMs to handle all the search traffics with acceptable cost and latency. However, the judgment of relevance is objective and non-personalized. This implies that the relevance scores for the same query-item pair should remain consistent for all users, unlike recommendation algorithms that are personalized for each individual. As long as we can obtain the relevance scores for the query-item pairs, we can perform online services for all users, which greatly reduces the volume of online requests. And for the certain query or item, their semantic information is relatively stable and does not vary much. Inspired by this, we come to a efficient and effective solution (i.e, online and offline collaborative service) with the affordable cost and latency. Figure~\ref{fig:OOCS} concretely illustrates online serving process of our proposed online and offline collaborative service when a certain query-item pair appears.

Specifically, we utilize LLMs (e.g., 10B parameters) to perform offline inference on daily search logs exploiting the daily prompts and update relevance scores stored in the database. The cost of offline inference is minimal, so we can choose LLMs with a larger number of parameters to enhance the accuracy of relevance score calculations. Meanwhile, we would deploy distilled smaller LLMs (e.g., 2B parameters) for online serving. Since the method of distillation is not the focus of this paper, we have omitted the introduction of the method here. With this model, it is possible to make online predictions for query-item pairs that do not have stored scores in the database. When a user enter a query in Alipay search, the service will firstly look up the database from the offline service and obtain relevance scores based query and candidate items. Once the relevance scores of the query-item pair can not be obtained from the database, the online service will be requested for predicting the relevance score timely. Note that approximately 95\% of the traffic will be handled by the offline service, while the remaining traffic will be handled by the online service. Through this method, we have achieved affordable efficiency and resource requirements for exploiting LLMs to handle relevance judgements of industrial scenarios.

\subsection{Baseline Models}
\begin{itemize}
    \item \textbf{DSSM} is a two-tower representation-based model. It encodes the embedding of a given query and item independently and computes the relevance score accordingly.
    
    \item \textbf{ReprBert} is a representation-based Bert model that utilizes novel interaction strategies to achieve a balance between representation interactions and model latency.
    
    \item \textbf{Bert} has achieved significant progress on NLP tasks as an interaction-based model. Here we concatenate the query and item as the input of the model.
    
    \item \textbf{MASM} leverages the historical behavior data to complete model pre-training as a weak-supervision signal with a newly proposed training objective.
    
    \item \textbf{TextGNN} extends the two-tower model with the complementary graph information from user historical behaviors.
    
    \item \textbf{AdsGNN} further proposes three aggregation methods for the user behavior graph from different perspectives.

    \item \textbf{BARL-ASe} proposes dual behavior-neighbors augmented relevance model with self-supervised learning. And it exploits LLMs to deal with long-tailed query-item pairs.

    \item \textbf{BLOOM} is a language model trained on 46 natural languages and 13 programming languages with the causal decoder architecture.

    \item \textbf{GLM} is a language model with both powerful natural language understanding and generation capabilities based on the prefix decoder.
\end{itemize}

\end{document}